\input amstex.tex
\magnification\magstep1
\documentstyle{amsppt}

%\redefine\Bbb{\bold}

\define\a{\alpha}

\redefine\b{\beta}
\define\g{\gamma}

\redefine\o{\omega}

\redefine\l{\lambda}
\redefine\L{\Lambda}

\define\gm{\bold g}

\define\<#1,#2>{\langle #1,#2\rangle}
\define\Tr{\text{tr}}
\define\dep(#1,#2){\text{det}_{#1}#2}
\define\norm(#1,#2){\parallel #1\parallel_{#2}}

\topmatter
\title CHIRAL ANOMALY ON A LATTICE \endtitle
\author Jouko Mickelsson \endauthor
\affil Theoretical Physics, Royal Institute of Technology, Stockholm,
S-10044 Sweden \endaffil
\endtopmatter

\document
\baselineskip 18pt
\NoBlackBoxes

ABSTRACT  A calculation of the chiral anomaly on a finite lattice without
fermion doubling is presented . The lattice gauge field is defined
in the spirit of noncommutative geometry. Standard formulas for the
continuum anomaly are obtained as a limit.

\vskip 0.5in

\define\Sl{\!\!\!\!\slash}
\define\ps{p\Sl}
\define\As{A\Sl}

\define\TR{\text{tr}}

1. THE FINITE LATTICE EFFECTIVE ACTION

\vskip 0.3in
Chiral fermions have been the Achilles' heel of lattice approximations
in QCD. In standard lattice method one uses a nearest neighbor approximation
to the derivative of a fermion field, the gauge field is represented by
the Wilson link variables for nearest neighbor lattice points. In this approach
one cannot avoid an unphysical doubling of fermion degrees of freedom.
In the present paper I want to propose an alternative lattice approximation which
uses not only the nearest neighbors but \it all points \rm on the finite
lattice in the construction of the (chiral) Dirac operator. This may sound
a bit clumsy, but actually we shall see that it is quite natural, and what is
most inportant, it leads to the correct continuum limit without a fermion
doubling.

We study chiral fermions in euclidean space in $2n$ dimensions. In the continuum
the fermions are functions in $\Bbb R^{2n}$ taking values in $\Bbb C^{2^n}
\otimes \Bbb C^N.$ The euclidean $\g$- matrices operate on the first factor
and the compact gauge group $G$ on the
second factor. The $\g$ matrices satisfy
$$\g_{\mu} \g_{\nu} +\g_{\nu} \g_{\mu} = 2\delta_{\mu \nu}.\tag1$$
The matrix $\g_{2n+1}= (-i)^n \g_1\g_2 \dots \g_{2n}$ anticommutes with the
other $\g$ matrices.
We define the chiral projections $P_{\pm}= \frac12(1\pm \g_{2n+1})$ and
consider the chiral Dirac operator
$$D_A = \g_{\mu}(-i\partial_{\mu}+ A_{\mu} P_+)= \ps + \As P_+\tag2$$
coupled to a vector potential $\As= \g_{\mu} A_{\mu}$ with $A_{\mu}$
functions on $\Bbb R^{2n}$ taking
values in the Lie algebra $\gm$ of the gauge group $G.$

On a finite lattice $L\subset \Bbb R^{2n}$ consisting of points $x= m\cdot
\Delta,$ where the lattice spacing $\Delta$ is some fixed positive real number
and $m=(m_1,m_2,
\dots, m_{2n})$ is a vector with integer components $n_{\mu}=-q,-q+1,\dots,
q,$
any function is a linear combination of the
Fourier modes $f_{p}(x)= \frac{1}{(2q)^{2n}} e^{i p\cdot x}$ with $p_{\mu}=
\frac{2\pi}{2q\Delta} k$
and $k=-q,-q+1,\dots, q.$  Thus the vector space of massless fermions on the
lattice $L$ has dimension $K=2^n\cdot N \cdot (2q)^{2n}.$ Note that at the end
points, corresponding to $k=\pm q,$ in the momentum space lattice the Fourier
modes are identical; we have periodicity with period $2q$ in the momentum
index $k.$ We shall later break this periodicity.

When studying the finite volume continuum limit one lets $q\to
\infty$ and $\Delta\to 0$ such that $\ell=2q\Delta$ is fixed. However, we
shall not insist on this but we keep the option open for an infinite volume
limit.

Now our approach departs from the usual setting. Normally, the Dirac operator
is written using the principle that partial derivatives on a lattice are
certain finite difference operators involving nearest neighbor points.
Here we define the free Dirac operator
through the Fourier transform $D_0 \mapsto \g_{\mu} p_{\mu}.$
The vector
potential $A=A_{\mu}\g_{\mu},$ thought of as an operator in the one-particle
Hilbert space, is defined as a $K\times K$ matrix with matrix elements
$$(\g_{\mu} A_{\mu})_{s p,s'p'} = (\hat A_{\mu}(p-p')\g_{\mu})_{ss'}\tag3$$
where $ss'$ denote both the spin and gauge indices and the ambiguity in the
Fourier transform $\hat A$ (when restricting the continuum potential to the
finite lattice) is resolved by requiring
that both $k,k'=-q,-q+1,\dots,
q$ for $p_{\mu}= k\cdot \frac{2\pi}{2q\Delta}.$  Note that in general the
Fourier transform \it is not periodic \rm and therefore the matrix elements
${\hat A}_{sp, s'p'}$ for $p,p'= \pm \frac{2\pi}{2q\Delta} q$ are all
independent. The number of lattice sites is $(2q+1)^{2n}$ instead of
$(2q)^{2n}.$  What we done above is simply that we have defined the lattice
operator $A$ (in momentum space) as a $K\times K$ submatrix of the
corresponding continuum operator by restricting the momentum indices to the
given range.

We shall generalize the above setting in the spirit of noncommutative geometry,
[C]. The reason for doing it here is purely technical: We do not need to think
all the time whether the matrix representing a vector potential really
represents a multiplication by a Lie algebra valued function.  A vector
potential in the generalized sense is any hermitean
$K\times K$ matrix of the form $A_{\mu}\g_{\mu}$ where the $A_{\mu}$'s are
hermitean matrices acting in $\Bbb C^{N\times (2q+1)^{2n}}.$ A gauge
transformation is then an unitary $K\times K$
matrix $g$ which commutes with the $\g_{\mu}$'s. The action of $g$ on a
potential $A$ is given as
$$\As'= g^{-1}\As g + g^{-1}[\ps,g].\tag4$$
The curvature form is
$$F_{\mu\nu} =[p_{\mu},A_{\nu}] -[p_{\nu},A_{\mu}] +[A_{\mu}, A_{\nu}].\tag5$$
It transforms as usual, $F_{\mu\nu}\mapsto g^{-1}F_{\mu\nu} g.$

We define the lattice effective action
$$  S_{eff}(A) = -\text{log}\, Z(A),\hskip 0.3in
Z(A)= {\det}_{ren}(\ps+\As P_+)\tag6$$
where $\det_{ren}$ is a renormalized determinant.  Even on a finite lattice we
want to use certain renormalized determinants in order that in the limit
$q,1/\Delta \to \infty$ the effective
action remains finite and leads to the continuum effective action.

In continuum the first renormalization ('vacuum subtraction') is to replace
the determinant $\det(\ps+\As P_+)$ by $\det(\frac{1}{\ps+i\epsilon}(\ps+\As_+
P_+)) \equiv\det(1+T),$ where $T$ is an operator of order $-1$ in momenta. The term
$i\epsilon$, for a real nonzero parameter $\epsilon,$ is introduced as an
infrared regularization. From now on any $\ps$ in the denominator stands for
$\ps+i\epsilon.$
Using the
formula $\text {log}\,\det = \TR \,\text{log}$ and the expansion of the
logarithm
$$\TR\, \text{log}(1+T)= \TR( T-\frac12 T^2 +\frac13 T^3 \dots)
\tag7$$
one localizes the potentially diverging terms
as the traces
of $T^k$ for $k=1,2,\dots,2n.$ The higher powers behave like $|p|^{-2n-a}$ as
$|p|\to\infty$
for some positive $a$. Assuming that the potential $A$ vanishes more rapidly
than $|x|^{-2n}$ at infinity, the trace of these terms is finite. In terms of
a momentum space cut-off $\L$ one can write an asymptotic expansion
$$\TR_{\L} (T- \frac12 T^2 +\dots -\frac{1}{2n} T^{2n})= \sum_{i\leq i_0}
\L^i a_i +\text{log}(\L/\L_0)a_{log}.\tag8$$
The scale fixing constant $\L_0$ is determined by the physical requirement that the
strength and location of the pole at $q=0$ of the boson propagator is not affected by the
loop diagrams. The renormalized trace is defined as the coefficient $a_0$ in the asymptotic
expansion.

It follows that we can define the effective action as proposed by Seiler [Se],
$$-\text{log} \,{\det}_k(1+T)+ \text{ a finite number of Feynman diagrams}$$
where
$$\text{log}\, {\det}_k (1+T)= \TR \left( \frac{(-1)^{k-1}}{k} T^k +\frac{(-1)^k}
{k+1} T^{k+1}+\dots\right)\tag9$$
with $k=2n+1.$
The finite set of (renormalized) Feynman diagrams comes from the renormalized
traces of $T^k$ for
$k \leq 2n.$ For example, when $2n=4$ the only diverging diagrams are the
vacuum polarization terms with at most four external gauge boson lines.

In the lattice the modified Fredholm determinant ${\det}_k(1+T)$ is defined
exactly as in the continuum case. The renormalized determinant is
$${\det}_{ren}(1+\frac{1}{\ps} \As P_+) ={\det}_{2n+1}(1+\frac{1}{\ps} \As P_+)
\cdot e^{\text{TR}(T +\dots +\frac{-1}{2n} T^{2n})}\tag10$$
where the renormalized trace TR is defined as follows.

\bf Case of $2n=2.$ \rm The continuum limit of $\TR_{\L}( \frac{1}{\ps} \As_+)$
vanishes by a simple parity argument. Thus no renormalization is needed for
this term. The next term $\TR_{\L}( T^2)$ is
potentially logarithmically diverging.
However, by the trace properties of products of $\g$ matrices and parity this
term is actually of order $-3$  and gives a finite trace. Thus the effective
action is completely determined by
$\text{det}_3(1+T)$ and the finite 1-loop diagram, $\text{TR}(T-\frac12 T^2)=
-\frac12 \TR_C T^2,$ where $\TR_C$ stands for the conditionally convergent
trace: one computes first the trace over spin and color indices, then integrates
over momentum variables with $|p| \leq \L,$ followed by the limit $\L\to\infty$
and the integration over $x.$

\bf Case of $2n=4.$ \rm The first term $\TR\, T$ vanishes  as in the previous
case. The next is the
1-loop diagram $\TR\, T^2$   with two external boson lines.  In the continuum
we must compute traces of operators which are composed of products of Green's
functions (the operators $1/(\ps+i\epsilon)$) and and of smooth functions $\As.$
They are examples of pseudodifferential operators. The algebraic manipulations
involving PSDO's are most conveniently performed using the symbol calculus.
First let us recall the basic rule of
symbol calculus: A pseudodifferential operator is represented by a smooth
function of coordinates and momenta, its symbol. The symbol $a*b$ for a product
of operators is computed from the symbols $a,b$ of the factors as
$$(a*b)(x,p)= \sum \frac{(-i)^{n_1+\dots n_d}}{n_1!n_2!\dots n_d!}
\frac{\partial^{n_1}\dots \partial^{n_d} a}{(\partial p_1)^{n_1} \dots
(\partial p_d)^{n_d}} \frac{\partial^{n_1}\dots \partial^{n_d} b}
{(\partial x_1)^{n_1} \dots (\partial x_d)^{n_d}}.$$
Now the first term in the expansion of the PSDO $(\frac{1}{\ps} \As P_+)^2$
leads to both
quadratic and logarithmic divergencies which are given by the integral
$$-\frac{1}{(2\pi)^4} \int d^4 x \int d^4 p  \frac{p^2}{(p^2+\epsilon^2)^2}
\TR A^2.$$
The lattice version of this is simply
$$-\frac{1}{(2\pi)^4} \TR (\frac{p^2}{(p^2+\epsilon^2)^2} A^2$$
In addition, there is are logarithmic divergencies
arising from the next terms in the expansion of $T^2$ as well as contributions
from the higher order terms
which involve the traces $\Tr(T^k)$ for $k=3,4.$ In the continuum case,
by a standard Feynman integral calculation carried out in [SABJ], one obtains ($\b$ is a
numerical constant) as the total logarithmic divergence
$$\b\, \text{log}(\L/\L_0) \int \TR (F^2 + F*F)$$
where $\L_o$ is a renormalization constant and $*F$ is the dual of the field
tensor. Actually, in our case the second term involving $F*F$ vanishes because
we are not considering instanton backgrounds, the vector potential is globally
defined and vanishes at $|x| \to \infty.$ A lattice version of this is
$$  \b   \TR \frac{1}{(p^2+\epsilon^2)^2} (F^2 +F*F) $$
modulo finite terms in the continuum limit.
Thus for $2n=4$  the effective action is given by $Z(A)=\det_5(1+T)
\times e^{\b(A)},$ where $\b(A)$ is the ordinary trace $\sum_{k=1}^4
\frac{-(-1)^k}{k}\TR(\frac{1}{\ps}\As_+)^k$ \it minus \rm the diverging terms
discussed above.

\bf General case. \rm In $2n$ dimensions there is a finite number of both
polynomially and logarithmically diverging terms.
A derivative in momentum space makes the diagram better converging. A derivative
in momentum space is associated with a differentiation in $x$ space of one
of the $A$'s in the expansion of the trace. Since for diverging diagrams
we can have only a finite number of differentiations, the coefficient of
$\L^k$ or of log$\L$ will be a finite differential polynomial in $A.$
The lattice renormalization is obtained by subtracting these diverging terms
from the naive effective action in such a way that the partial derivatives
$\frac{\partial}{\partial x_{\mu}}$ are replaced by  the multiplication
operators $ip_{\mu}.$

\vskip 0.3in
2. THE ANOMALY

\vskip 0.3in
Next we shall compute the gauge variation of the (lattice) effective action.
For that we need some properties of the generalized Fredholm determinants.
According to our definition, [S],
$${\det}_k(1+X) =\det[(1+X)e^{\b_k(X)}],\tag11$$
where
$$\b_k(X)= \sum_{i=1}^{k-1} \frac{(-1)^i}{i} X^i.\tag12$$
If all the traces of powers of $X$ are finite we can write
$${\det}_k(1+X) = \det(1+X) \cdot e^{\TR \b_k(X)}.\tag13$$
The generalized determinants have a multiplicative anomaly,
$${\det}_k[(1+X)(1+Y)] = {\det}_k(1+X) \cdot {\det}_k(1+Y)\cdot e^{\g_k(X,Y)},
\tag14$$
where
$$\g_k(X,Y) = \TR [\b_k(X+Y+XY)-\b_k(X) -\b_k(Y)].\tag15$$
The first nonzero multiplicative anomaly is $\g_2(X,Y)= -\TR(XY),$ the next is
$\g_3(X,Y)= \TR(X^2 Y+Y^2 X +\frac12 (XY)^2).$
We shall also need the derivative $H_k=\g_k^{(1)}$ with respect to the first
argument at $(X,Y)=(0,Y).$ Its value for a variation $\delta X$ is easily
computed to be
$$H_k(Y) ( \delta X)= \TR (Y^{k-1} \delta X).\tag16$$

The gauge variation of $S_{eff}(A)$ can be computed as follows.
Modulo a variation of a finite polynomial in $A,$ the anomaly is given by
the gauge variation of the functional $Z_k(A)={\det}_k (1+\frac{1}{\ps} \As_+),$
$\As_+= \As P_+.$
Denote $g_-= g P_+ +P_-$ and $g_+= gP_+  +P_-.$
We obtain
$$\align Z_k(A^g)&= Z_k(g^{-1} Ag +g^{-1} dg)\\
&={\det}_k(\frac{1}{\ps} g_- (\ps+\As_+) g_+^{-1})\\
&={\det}_k(g_+^{-1} \frac{1}{\ps} g_- \ps (1+\frac{1}{\ps} \As_+))\\
&={\det}_k( g_+^{-1} \frac{1}{\ps} g_- \ps) \cdot {\det}_k(1+\frac{1}{\ps}
\As_+) \cdot e^{\g_k(g^{-1} \frac{1}{\ps} [\ps,g] P_+, \frac{1}{\ps} \As_+)}.
\tag17\endalign$$

For infinitesimal gauge variations we have
$$\o(X;A)=\delta_X  S_k(A)= \frac{d}{dt} S_k(A(t))|_{t=0}
= (D_1\g_k(0,\frac{1}{\ps}\As_+)) ( \frac{1}{\ps}[\ps,X] P_+),\tag18$$
where $D_1$ is the derivative with respect to the first argument and
$A(t)= A^{g(t)}$ for $g(t)= e^{tX}.$ By (16),
$$\o(X;A)= \TR (\frac{1}{\ps} [\ps,X] (\frac{1}{\ps} \As_+)^{k-1}).\tag19$$

\bf Example $2n=2$ \rm Because of the accidental property of the two dimensional
effective action, $(\frac{1}{\ps}\As P_+)^2$ is conditionally convergent,
we may use the determinant ${\det}_2$ (instead of the more complicated
${\det}_3$). The anomaly is then computed from the multiplicative anomaly
$\gamma_2.$ Both on the lattice and the continuum the anomaly is
$$\o(A;X)= \TR\left( \frac{1}{\ps+i\epsilon}[\ps,X]P_+\frac{1}{\ps+i\epsilon}
\As P_+ \right).\tag20$$
In the continuum a simple computation leads to
$$\o(A;X) = \frac{i}{4\pi} \int d^2 x \TR A_{\mu} \partial_{\nu} X\epsilon^{
\mu\nu} + \frac{1}{2\pi} \int d^2 x \TR A_{\mu} \partial_{\mu} X.\tag21$$
The second term is a trivial cocycle, it is the gauge variation of the local
functional
$$\frac{1}{4\pi}\int d^2x \TR A^2.$$
The first term is the nontrivial part of the anomaly. On the lattice the local
formula does not make sense, but we still have the trace formula (20), which
in the continuum limit leads to an integration in momentum and configuration
space, giving the formula (21).

\bf The general case. \rm The nontrivial part of the anomaly comes from the
difference
$$\theta_k(A)=\text{log}\,\det(1+T) - \text{log}\, {\det}_k(1+T).\tag22$$
This follows form the fact that the unrenormalized determinant is perfectly
gauge symmetry; the gauge symmetry is spoiled by the renormalization. In the
continuum the difference above is infinite, but we may use this formula on the
finite lattice, giving
$$\o(A;X) = \delta_X \theta_k(A)=\delta_X \sum_{j=1}^{k-1}\TR \frac{(-1)^{j-1}}
{j} T^j.\tag23$$
Inserting $T=\frac{1}{\ps}\As_+$ and $\delta_X \As = [\As,X]+[\ps,X]$ in (23)
we obtain once more the form (19) of the anomaly.

In the case of periodic boundary conditions in the continuum version (or
more generally, with an enough rapid decrease of the gauge fields at $|x| \to
\infty$) we can prove that  (23) has a continuum limit as the momentum space
lattice size is increased (the cut-off $|p| \leq \L$ is removed). This follows
from a simple H\"older inequality argument: If $W$ is any operator such that
$|W|^n$ is a trace-class operator, that is $W\in L_n,$ and $W_i$ is a sequence
of operators in $L_n$ converging to $W$ with respect to the $L_n$ norm
then lim$\, \TR (W_i)^n = \TR W^n.$ Now $W=T$ and the $W_i$'s are $i\times i$
lattice approximations to the continuum operator $T,$  lim$\,|W-W_i|_n=0.$

The operators under the trace
map in (23) become conditionally trace-class for $k\geq 2n$, [LM1],
and the trace of the infinite dimensional matrices (continuum limit) is given by a
local formula, the eq. (21) in the case $2n=2.$

\vskip 0.3in

3. HAMILTONIAN FORMULATION: ANOMALY OF THE CURRENT ALGEBRA

\vskip 0.3in
The same finite lattice approximation can be used also in the real time
hamiltonian formulation for fermions in background gauge fields.
The hamilton operator $D_A$ in $2n-1$ dimensions is defined exactly in the
same way as the $2n$ Dirac operator earlier,
$$D_A = \g_0\g_k (p_k +A_k)=\a_k(p_k +A_k),\tag24$$
sum over $k=1,2,\dots,2n-1.$ The dimension of the space lattice is now
$K=2^n \cdot N\cdot (2q+1)^{2n-1}$, in case of Dirac fermions; for Weyl fermions the
hamiltonian above must be multiplied by the chiral projection $P_+$ and the
dimension of the projected subspace is $K=2^{n-1}\cdot
N\cdot (2q+1)^{2n-1}.$

We shall consider the gauge currents for massless fermions. The left and
right components of
fermions decouple and we may restrict to (left-handed) Weyl fermions. The
current algebra for Dirac fermions becomes just the direct sum of left and
right current algebras.

In continuum the current algebra is anomalous. There are Schwinger terms which
in the case $2n-1=1$ can be written as
$$[\rho(X),\rho(Y)]= \rho([X,Y]) +  \frac{i}{2\pi}\int_M \TR X(x) Y'(x) dx,
\tag25$$
where the integral is over the one dimensional space $M.$ When $M$ is a unit
circle (25) gives an affine Kac-Moody algebra. Here $\rho(X)$ denotes
the charge density integrated with a smooth Lie algebra valued test function
$X,$
$$\rho(X)= \int_M \rho_k(x) X_k(x) dx,$$
where $k$ is a Lie algebra index, $k=1,2,\dots,\text{dim}\,\gm.$ The trace under
the integral sign in (25) refers to the representation of the gauge group
acting on fermion components. In the case $2n-1=3$ one has, [M1], [F-Sh],
$$ [\rho(X),\rho(Y)]= \rho([X,Y]) + \frac{1}{24\pi^2}\int_M \TR A [dX,dY],
\tag26$$
where the 3-form under the integral is defined as an exterior product of the
1-forms $A,dX,dY.$

There are alternative, but equivalent, formulas for the Schwinger terms. Equivalent
means again that the difference between the Schwinger terms is a gauge variation,
of the type $\Cal L_X \theta(A;Y) -\Cal L_Y\theta(A;X) -\theta(A;[X,Y])$ for some function
$\theta$ of $A$ and $X,$ linear in the latter argument.
In the case  $2n-1=1$ one has in fact an exact formula, [L],
$$c(X,Y)= \frac14 \TR \epsilon[\epsilon,X][\epsilon,Y],\tag27$$
where $\epsilon$ is the sign of the free 1-particle hamilton operator.
In three space dimensions we have, [MR],[LM1],
$$c(A;X,Y)= \frac18 \TR_C (\epsilon-F)[[\epsilon,X],[\epsilon,Y]],\tag28$$
where now $F$ is the sign of the 1-particle hamiltonian $D_A$ in the external
gauge field $A.$  The conditional trace is defined as $\TR_C K= \frac12 \TR
(\epsilon K+K\epsilon).$

These formulas, and the corresponding formulas in higher dimensions, can be
directly translated to the lattice. When defining the sign operators $F,\epsilon$
one should be careful in order to have the right continuum limit. We have to
split (somewhat artificially) the energy spectrum even in the finite case
to positive and negative parts. As before we define the momentum
components as $p_i= \frac{2\pi}{2q\Delta} k$ with $k=0,\pm 1, \pm 2,\dots,\pm q$
for some positive integer $q$. The energy eigenvalues for the free hamiltonian
become then $E= \pm (p_i p_i)^{1/2}$ and so the spectrum is symmetric around
zero.

In the case of periodic boundary conditions (in $x$ space) in the continuum
theory it is simple to prove that one obtains the right continuum limit
as the size of the momentum space lattice is increased, that is, when the
cut-off $|E| \leq \L$ is removed. This follows again from a standard H\"older
inequality argument. The trace in (28) is convergent, it is known that the
operators $[\epsilon,X],[\epsilon,Y]$ belong to the Schatten ideal $L_4$
of operators $T$ such that $|T|^4$ is trace-class; furthermore, the diagonal
blocks $\epsilon(\epsilon-F) +(\epsilon -F) \epsilon$ of $\epsilon -F$
(the only part of the operator which contributes to (28)) are Hilbert-Schmidt.
All these operators can be approximated by finite-dimensional matrices
(in the appropriate $L_k$ norms) and therefore the trace of the product is a
limit of traces of finite-dimensional (cut-off) matrices.

We have discussed above the abstract commutation relations of the current
algebra but we have said nothing about an operator realization of the
currents. In $1+1$ dimensions the theory is well understood; a physically
acceptable realization is obtained using highest weight representations of
affine algberas. An important example is the basic representation which is
a representation in a fermionic Fock space. This construction has been
generalized to the $3+1$ dimensional case (and the method works in higher
dimensions), [M2]. That representation is also suitable for a lattice
approximation, as will be breafly explained below.

The basic idea is to define a continuous family $T_A$ of unitary conjugations
in the one-particle fermionic Hilbert space such that the off-diagonal
blocks (with respect to the energy polarization $\epsilon$) of currents
are reduced such that the resulting (unitarily equivalent) Gauss law
generators can be quantized by canonical methods. More precisely, we have to
require that
$$[\epsilon, \theta(X;A)] \text{ is Hilbert-Schmidt }, \tag29$$
where $X$ is any infinitesimal gauge transformation and
$$\theta(X;A)=  {T_A}^{-1} (\delta_X +X) T_A -\delta_X.\tag30$$
Note that the modified Gauss law generators $\tilde G_X= {T_A}^{-1} G_X T_A,$
with $G_X= \delta_X +X,$ automatically satisfy the same commutation relations as the
generators $X.$  The construction of the operators $T_A$ is best understood
through an (asymptotic) expansion in powers of the inverse momenta $1/\ps.$
In three space dimensions the first terms of the symbol are, [M2],
$$T_A= 1 - \frac14 \frac{1}{p}[\ps,\As]\frac{1}{p}+\dots\tag31$$
and the resulting gauge currents
$$\theta(X;A)= X +\frac{i}{4} \frac{1}{p}[\ps, \sigma_{\mu}]\partial_{\mu} X
\frac{1}{p} +   \dots,\tag32$$
where we have used the hermitean Pauli matrices $\sigma_{\mu}$ as the
3-space Dirac matrices, $\ps=\sigma_{\mu} p_{\mu}.$ In all formulas it is
implicitely assumed that an infrared regularization $1/|p| \mapsto 1/(|p|+
\l)$ is performed.
The terms which are of order strictly lower than $-1$ in momenta are
Hilbert-Schmidt and therefore not critical for the current renormalization.
With the renormalized current operators in hand, one can compute the quantum
commutation relations in a straight-forward way. The resulting Schwinger term
was calculated in [M2] and found to be equivalent
with (26).

These formulas can be translated immediately to the lattice. One replaces
the derivatives $-\partial_{\mu} X$ by $[p_{\mu},X]$ and the momentum symbols
in the PSDO's become multiplication operators by the discrete momentum
variables.

\vskip 0.3in
\it Acknowledgement \rm I wish to thank S. Rajeev for bringing
to my attention  the Schwinger formula for logarithmic divergence and
Edwin Langmann for enjoyable discussions and for suggesting improvements in the
manuscript.

\vskip 0.3in
\bf References \rm

[C] A. Connes: \it Noncommutative Geometry. \rm Academic Press (1994)

[F-Sh] L. Faddeev and S. Shatasvili, Theor. Math. Phys. \bf 60, \rm
770 (1984)

[LM1] E. Langmann and J. Mickelsson, Phys. Lett. \bf B338,  \rm 241 (1994)

[LM2] E. Langmann and J. Mickelsson, Lett. Math. Phys. \bf 36, \rm 45 (1996)

[L] L.-E. Lundberg, Commun. Math. Phys. \bf 50, \rm 103 (1976)

[M1] J. Mickelsson, Lett. Math. Phys. \bf 7, \rm 45 (1983);
Commun. Math. Phys. \bf 97, \rm 361 (1985)

[M2] J. Mickelsson, in \it Constraint Theory and Quantization Methods, \rm
ed. by Colomo, Lusanna, and Marmo, World Scientific (1994); also in \it
Integrable Models and Strings, \rm ed. by Alekseev et al., Springer LNP 436
(1994)

[MR] J. Mickelsson and S. Rajeev, Commun. Math. Phys. \bf 116, \rm 365 (1988)

[SABJ] J. Schwinger, Phys. Rev. \bf 82, \rm 664 (1951); S. Adler, Phys. Rev.
\bf 177, \rm 2426 (1969); J.S. Bell and R. Jackiw, Nuovo Cimento \bf 60A, \rm
47 (1969)

[Se] E. Seiler, Phys. Rev. \bf D 22, \rm 2412 (1980); Phys. Rev. \bf D 25, \rm
2177 (1982)

[S] B. Simon: \it Trace Ideals and Their Applications. \rm Cambridge
University Press (1979)

\enddocument